\documentclass[final]{svjour2}
\usepackage[pdftex]{graphicx}
\usepackage{sidecap}
\usepackage{graphicx}
\usepackage{rotating}
\usepackage{amsmath}
\usepackage{amssymb}
\usepackage{mathptmx}
\usepackage{amssymb}
\usepackage{epstopdf}
\usepackage{mathrsfs}
\usepackage{sidecap}
\usepackage{subfigure}
\usepackage{comment}
\usepackage{float}
\usepackage{graphicx,pb-diagram,mathrsfs}
\usepackage{eso-pic}
\usepackage {graphicx}
\usepackage[numbers]{natbib}
\makeatletter
\journalname{Journal of Low Temperature Physics}
\bibpunct{}{}{,}{s}{}{,}

\begin{document}
\newcommand{\vv}{\mathbf}
\title{Scattering of Line-Ring Vortices in a Superfluid}

\author{Alberto Villois$^1$ \and Hayder Salman$^1$  \and Davide Proment$^1$}

\institute{1:School of Mathematics, University of East Anglia, Norwich Research Park, NR4 7TJ Norwich, UK\\
%Tel.:\\ Fax:\\
\email{H.Salman@uea.ac.uk}
}

\date{14.10.2014}

\maketitle


\begin{abstract}

We study the scattering of vortex rings by a superfluid line vortex using the Gross-Pitaevskii equation in a parameter regime where a hydrodynamic description based on a vortex filament approximation is applicable. By using a vortex extraction algorithm, we are able to track the location of the vortex ring as a function of time. Using this, we show that the scattering of the vortex ring in our Gross-Pitaevskii simulations is well captured by the local induction approximation of a vortex filament model for a wide range of impact parameters. 
%This provides further justification for analysing vortex ring and line interactions at intermediate scales of turbulence using the LIA. 
%At such scales, the vortex lines appear well-separated while the excitations on individual lines remain sufficiently large such that a nonlinear Kelvin-wave cascade has not been established at that stage. 
The scattering of a vortex ring by a line vortex is characterised by the initial offset of the centre of the ring from the axis of the vortex. We find that a strong asymmetry exists in the scattering of a ring as a function of this initial scattering parameter.\\
% that arises as a consequence of topological constraints imposed at the point of reconnections between the ring and the line vortex.
%We derive simple laws that determines how the vortex ring scatters based on constraints of conservation of momentum, energy, topology, angular momentum.
%\textcolor{red}{Need to shorten abstract.}

\end{abstract}

\section{Introduction}
Quantized vortex rings in superfluids constitute one of the most fundamental localised topological excitations in the study of quantum turbulence. In recent years, such vortex rings have attracted much attention as they continue to provide a key mechanism in understanding the transition to quantum turbulence\cite{Fujiyama2010}. At the same time, they can serve as a vital mechanism in understanding how different regimes of turbulence can coexist on disparate length and time scales. When forced at large scales, quantum turbulence can give rise to a Kolmogorov like inertial range in the energy spectrum. It has been proposed that this gives way to the Kelvin wave cascade at higher wave numbers for liquid $^4$He within the ultra-low temperature regime\cite{Walmsley08,Svistunov2009}. Within the cross-over range of scales, the emission of vortex rings from individual quantized vortex filaments acts to assist in the transfer of energy from large quasi-classical motion to the small Kelvin-wave cascade. Indeed, without such a mechanism, a bottleneck can form within the energy cascade\cite{Lvov2007}. However, once these rings are emitted, their subsequent role in the direct energy cascade of superfluid turbulence remains unclear. This is particularly so since no measurements of the characteristics of turbulence within the cross-over range has thus far been made. The possible scenarios are that the emitted vortex rings can either pass through the tangle with little interaction or they could be reabsorbed. This question of the transparency of the tangle to the emission of vortex rings remains an open and not well understood problem in superfluid turbulence.

The important role that vortex rings occupy in the study of quantum turbulence has prompted detailed experimental studies of these excitations in both $^4$He and $^3$He
\cite{Bradley2005,Walmsley2014}. This has revealed that the number density of vortex rings strongly determines the likelihood of neighbouring vortex rings to reconnect. This subsequently influences their time of flight. Such studies can help uncover the nature of turbulence within the cross-over range of scales. However, a better understanding of how vortices interact and scatter with vortex lines is also important in order to establish  a clear understanding  of the role of vortex rings within this range of scales. In this paper, we will present a detailed numerical study of the scattering of a single vortex ring with a straight superfluid vortex. This scenario is in contrast to the scattering between vortex rings\cite{Caplan2014} and can be viewed as a paradigm of how vortex rings interact with vortex lines or bundles \cite{Sultan08} within fully developed quantum turbulence. Our approach relies primarily on the mean field theory of the Gross-Pitaevskii (GP) equation. However, to relate this to the hydrodynamic interpretation of quantum turbulence, we also interpret our results using the Biot-Savart law and its simpler approximate form given by the local induction approximation (LIA). To accomplish this, we develop diagnostics to extract the location of topological excitations from our mean-field model simulations that allows us to track the evolution of key quantities including vortex line length that provide a direct measure of key attributes associated with these topological defects.
 
\section{Mathematical Models}
We will model the interaction of quantized vortices within the mean-field theory of a superfluid which is governed by
\begin{equation}\label{GP}
 i\hbar{\partial_t \psi}= \frac{-\hbar^2}{2m}\nabla^2\psi+g|\psi|^2\psi.
 \end{equation}
While this equation provides a good description of the time evolution of a homogeneous weakly interacting dilute Bose gas for temperatures close to absolute zero, we will use it as a good qualitative model that can reproduce the essential hydrodynamics on sufficiently large length scales.
In Eq.\ \eqref{GP}, $\psi$ is the condensate wave-function for $N$ bosons of mass $m$, $\hbar$ is the reduced Planck constant, and $g=4\pi\hbar^2a/m$ is the interaction strength arising from binary interactions where $a$ is the $s$-wave scattering length.
Equation~\eqref{GP} is derivable from a Hamiltonian such that it satisfies the key conservation laws of mass
 \begin{equation}
 M=m\int|\psi|^2d^3\vv{x},
\end{equation}
linear momentum
\begin{equation}
 \vv{P}=\frac{\hbar}{2i}\int[\psi^*\nabla\psi-\psi\nabla\psi^*]d^3\vv{x},
\end{equation}
and energy
\begin{equation}\label{E}
 E=\frac{\hbar^2}{2m} \int \left( |\nabla\psi|^2 + \frac{g}{2} |\psi|^4 \right) d^3\vv{x}.
\end{equation}
It is useful to recast the above equation into hydrodynamic form by introducing the Madelung transformation
\begin{equation}\label{mad}
\psi=Re^{iS}.
\end{equation}
Defining the mass density, $\rho=mR^2$, and the local velocity vector, $\vv{v}=\frac{\hbar}{m}\nabla S$,
it is possible to show that the Gross-Pitaevskii equation transforms to
\begin{eqnarray}\label{eqcont}
&&  \frac{\partial \rho}{\partial t}+\nabla\cdot(\rho\vv{v})=0, \\
\label{EulerQ}
&& \frac{\partial \vv{v}}{\partial t}+\vv{v}\cdot \nabla\vv{v}+\frac{g}{m^2}\nabla\rho-\frac{\hbar^2}{2m^2}\nabla\left[\frac{\nabla^2\sqrt{\rho}}{\sqrt{\rho}}\right]=0,
\end{eqnarray}
which describes the flow of an irrotational and inviscid fluid. Even though the fluid is irrotational, the superfluid can support vortex filaments with vorticity concentrated according to $\boldsymbol{\omega}=\nabla\times\vv{v} = \kappa\delta(\vv{x}-\vv{s})$
where $\vv{s}=\vv{s}(\zeta_0)$ describes the position of the vortex filament that is parametrised by the arclength $\zeta_0$ at some initial time $t_0$. Here, $\kappa=n\hbar/m$ is the circulation around a superfluid vortex which is quantized since $n\in\mathbb{Z}$. Therefore, vortices within the GP model are delineated by curves $\vv{s}(t,\zeta_0)$ that can move. In contrast to a classical inviscid fluid described by the Euler equations, the GP model allows the topology of the vorticity field to change. Such changes in topology occur during reconnections and are an essential process in the decay of quantum turbulence.

To allow further analogies to be drawn between the GP equation and the hydrodynamic description of a classical fluid, we can decompose the energy appearing in Eq.\ (\ref{E}) into three contributions corresponding to a kinetic energy, potential energy, and quantum energy. These are given, respectively, by
\begin{eqnarray}
E_{kin} = \frac{1}{2}\int \rho|\vv{v}|^2d^3\vv{x}, \;\;\;\;\;\;\;
E_{int} = \frac{g}{2}\int\rho^2d^3\vv{x}, \;\;\;\;\;\;\;
E_{qu} = \frac{\hbar^2}{2m}\int|\nabla\sqrt{\rho}|^2d^3\vv{x}.
\label{eqn_En_cont}
\end{eqnarray}
Equations \eqref{eqcont} and \eqref{EulerQ} show that our system is in fact a compressible fluid.
The kinetic energy can, therefore,  be decomposed into an incompressible and a compressible component. Using the definition given by Nore {\em et al.}\cite{Brachet1997}, we introduce a Helmholtz-Hodge decomposition on the velocity weighted by the square root of the superfluid density such that
\begin{eqnarray}\label{Enkdec}
&& \sqrt{\rho}\vv{v}=(\sqrt{\rho}\vv{v})^i+(\sqrt{\rho}\vv{v})^c, \\
&& \nabla \cdot (\sqrt{\rho}\vv{v})^i = 0, \;\;\;\;\;\;\; \nabla \times \nabla (\sqrt{\rho}\vv{v})^c = 0.
\end{eqnarray}
It has become standard in the literature to interpret the incompressible and compressible energies given by
\begin{equation}
E_{kin}^i=\frac{1}{2}\int \left| (\sqrt{\rho}\vv{v})^i \right|^2 d^3\vv{x}, \;\;\;\;\;\;\;\;\;
E_{kin}^c=\frac{1}{2}\int \left| (\sqrt{\rho}\vv{v})^c \right|^2 d^3\vv{x},
\label{eqn_Ek_inc_comp}
\end{equation}
 as the energies associated with the vortex lines, and the sound excitations, respectively.
 
For a superfluid, the healing length $\xi$ characterises the size of the core of a vortex which in terms of quantities appearing in Eq.\ (\ref{GP}) can be expressed as $\xi = \hbar/\sqrt{2g\rho_{\infty}}$, where $\rho_{\infty}$ is the density of the fluid in the far field. For well separated vortices that have characteristic length scales $l \gg \xi$, variations due to density within the core can be neglected. In this case, the incompressible component of the kinetic energy dominates over the compressible kinetic energy and quantum energy and Eqs.\ \eqref{eqcont}-\eqref{EulerQ} simplify to the incompressible form of the fluid equations given by
\begin{eqnarray}
\nabla\cdot\vv{v} &=& 0, \\
 \frac{\partial \vv{v}}{\partial t}+(\vv{v}\cdot\nabla)\vv{v} &=& -\frac{1}{\rho}\nabla\rho.
\end{eqnarray}
On these length scales, our flow is considered to apply to a non-simply connected region that is threaded by vortex filaments. Therefore, the flow can still have a non-trivial circulation around each vortex filament. From the Biot-Savart law \cite{Donnelly1991}, we have that the velocity induced by a vortex filament is given by
\begin{equation}\label{BS}
\vv{v}=\frac{\kappa}{4\pi}\int_{\mathcal{C}(t)}\frac{d\vv{s}\times (\vv{x}-\vv{s})}{|\vv{x}-\vv{s}|},
\end{equation}
where $\vv{x}$ is position vector for any point in the fluid while $\vv{s}$ is a point of the curve $\mathcal{C}(t)$ where the vortex line lies.
According to the Kelvin-Helmholtz theorem\cite{Saffman1992}, we can also write the equation of motion for the vortex line as
\begin{equation}
\dot{\vv{s}}=\frac{\kappa}{4\pi}\int_{\mathcal{C}(t)}\frac{d\vv{s_0}\times (\vv{s}-\vv{s_0})}{|\vv{s}-\vv{s_0}|}.
\end{equation}
In general, the motion of a point lying on the vortex line depends on its instantaneous configuration. However, there are cases where one can consider the motion to be determined by only the local radius of curvature. This leads to 
the LIA\cite{Donnelly1991} 
\begin{equation}
\dot{\vv{s}}=\frac{\kappa}{4\pi}\ln{(1/ka_o)}\vv{s'}\times\vv{s''},
\end{equation}
where $\vv{s}'$ ans $\vv{s}''$ are respectively the first and second derivative of the curve with respect to the arclength, $k$ is a characteristic wavenumber of the vortex at the point $\vv{s}$, and $\xi \ll a_o \ll 2\pi/k$ is an intermediate cut-off scale. We remark that such a model is a completely integrable system for which the length of a vortex line is a constant of motion.

\section{Numerical Method}
\begin{SCfigure}
%\centering
\caption{The initial configuration used to study the scattering between a line vortex and a vortex ring. The ring, having initial radius $R$, lies in the $y$-$z$ plane while the line is aligned with the $z$-coordinate direction and passes through the origin $(0,0)$ (color figure online).}
\includegraphics[scale=0.3]{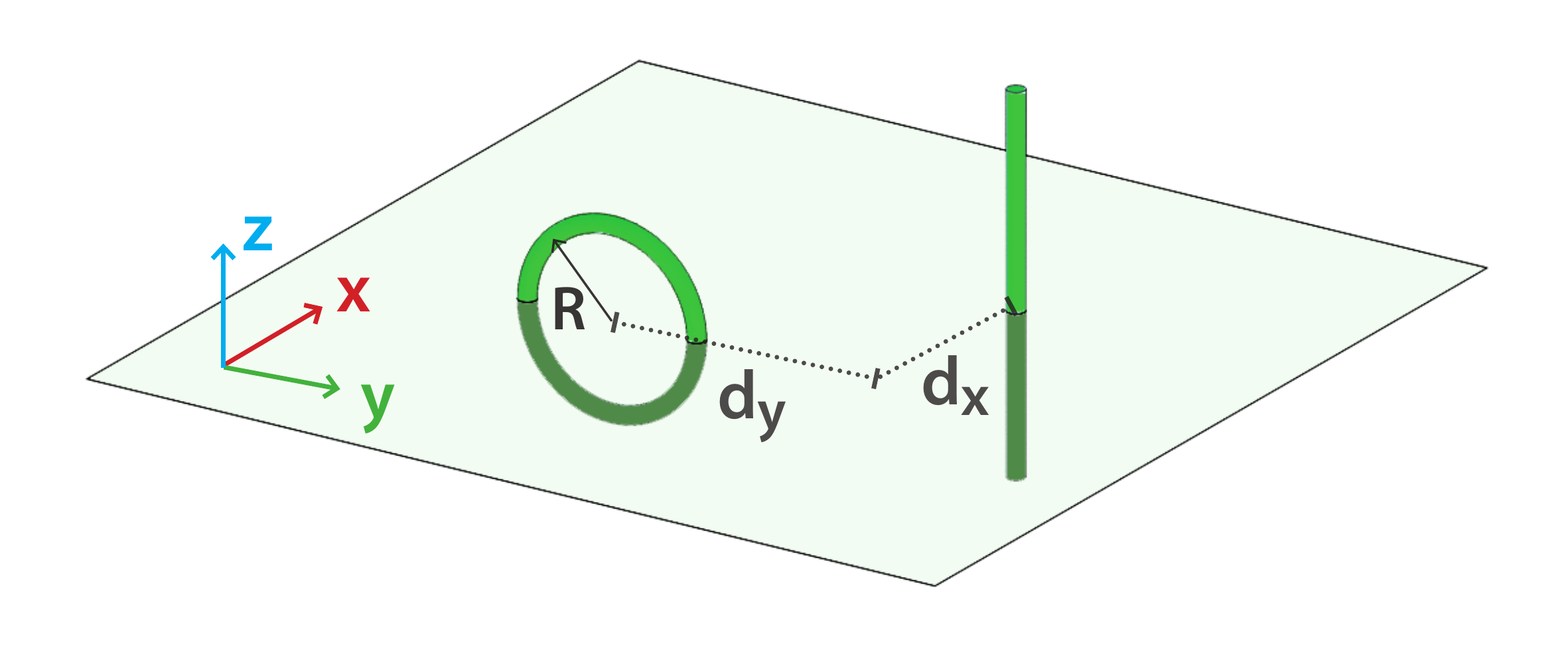}
\label{sketch}
\end{SCfigure}

%all present
In this work, we numerically solve a dimensionless form of the GP equation. Using the transformations $\psi\rightarrow \sqrt{\rho_{\infty}/m}\psi$, $x\rightarrow\xi x$ and $ t\rightarrow \tau t$, where $\xi=\hbar/\sqrt{2}mc$ is the healing length, $c=\sqrt{g \rho_{\infty}/m^2}$ is the speed of sound, and $\tau=\xi/\sqrt{2}c$, Eq.\ ~\eqref{GP} becomes
\begin{equation}\label{GPadim}
 i\partial_t\psi = \frac{-1}{2}\nabla^2\psi+\frac{1}{2}|\psi|^2\psi.
\end{equation}
The initial conditions we use for Eq.\ \eqref{GPadim} always correspond to a single straight line vortex and a vortex ring that is perfectly circular.
We make the hypothesis that the ring is small compared to the line by assuming that the radius, $R$, of the ring is much smaller than the distance, $d_x$, between the ring and the line vortex. Due to the size of the computational domain that extends over a length $L=128$ in each direction, we are restricted to set $R/d_x=1/6$. The boundary conditions are taken to be periodic along the $z$-coordinate direction which coincides with the axis of the straight line vortex, and reflecting in the other two directions corresponding to the $x$-$y$ plane. Figure \ref{sketch} provides an illustration of a typical initial condition.

The complex wave-function $\psi$ is discretised in a cubic box having a uniform grid of $256^3$ points. In order to accurately resolve the vortex core, we set the grid resolution to $\Delta x=\Delta y=\Delta z= 0.5$ so that $L=128$ as required.
The radius of the ring is set to $R=8$ and its initial distance from the line is set to $d_x=48$ to minimize the influence of the reflective boundaries and the line vortex on the ring. The line vortex is centred at the origin located at the centre of the computational domain while the vortex ring is located at $(-d_x, d_y, 0)$ and lies in the $y$-$z$ plane.
The initial condition is then obtained by setting $\psi(\mathbf{x}, t=0) = \psi_{line}(x,y,z) \times \psi_{Ring}(x,y,z)$ where
\begin{equation}
\psi_{Line}(x,y,z)=f(r)e^{i n \theta},
\label{eq:wf_line}
\end{equation}
with $r=\sqrt{x^2+y^2}$ and $\theta=\arctan(y/x)$.
This irrotational vortex has quantized circulation characterized by the discrete winding number $n$. The radial density profile $f(r)$ can be obtained from a Pad\'{e} approximation as described by Berloff\cite{Berloff2004}.
The solution for a vortex ring with radius $R$ moving along the $x$-direction is given by\cite{Proment}
\begin{equation}
\psi_{Ring}(x,y,z)=f(\eta_1)f(\eta_2)e^{i(\arctan{((\sqrt{z^2+y^2}-R)/x)}-\arctan{((\sqrt{z^2+y^2}+R)/x)})},
\label{eq:wf_ring}
\end{equation} 
where $\eta_1=\sqrt{(\sqrt{z^2+y^2}-R)^2+x^2}$ and $\eta_2=\sqrt{(\sqrt{z^2+y^2}+R)^2+x^2}$.

The non-dimensional GP equation \eqref{GPadim} is integrated forward in time using a standard operator {\it split-step} method. 
Defining the linear and nonlinear operators appearing in Eq.\ (\ref{GPadim}) 
as $\mathscr{L}=1/2\nabla^2$ and $\mathscr{N}=-1/2|\psi|^2$ respectively, we can write for a sufficiently small time step $\Delta t$
\begin{equation}\label{SplitStep}
\psi(\vv{x}, t+\Delta t)=e^{i \Delta t(\mathscr{N}+\mathscr{L})}\psi(\vv{x},t)\approx e^{i \frac{\Delta t}{2} (\mathscr{N})} \, e^{i \Delta t(\mathscr{L})} \, e^{i \frac{\Delta t}{2} (\mathscr{N})} \, \psi(\vv{x},t) \, .
\end{equation}
This approximation has an error of order $\mathcal{O}({\Delta t}^2)$.
The integration of the nonlinear operator can be easily computed in physical space using 
\begin{equation}\label{Non-Lin}
\psi(\vv{x},t)=\psi(\vv{x},t^*) \, e^{-i |\psi|^2 (t-t^*)/2}.
\end{equation}
On the other hand the integration of the linear operator part is performed in Fourier space as
\begin{equation}\label{Lin}
\tilde\psi(\vv{k},t)=\tilde\psi(\vv{k},t^*) \, e^{-i \omega(\vv{k}) (t-t^*)},
\end{equation}
where $\tilde{(\cdot)}$ denotes the Fourier-cosine transformed quantity and $\omega(\vv{k})=|\vv{k}|^2/2 $ is the angular frequency of the Fourier mode $\vv{k}$.
This is accomplished by decomposing the field $\psi$ using discrete cosine transforms in the $ x, y $ directions and the discrete Fourier transform in the $z$ direction in order to satisfy the reflective and periodic boundary conditions in the respective coordinate directions. 
In all our simulations, we set $\Delta t=0.02$, such that it is much smaller than the fastest linear period $ T=2\pi/\omega $ of the system.

\section{Results}
\begin{figure}
\centering
%\hspace{-0.5cm}
 \begin{minipage}[b]{1.0\textwidth}
    \vspace{0pt}
    \centering
    \subfigure[]{
%      \label{fig:mini:subfig:a}
%      \epsfig{file=figs/Fig_VLdens_comp1.eps,scale=0.014,angle=0.0}}
      \includegraphics[scale=0.5]{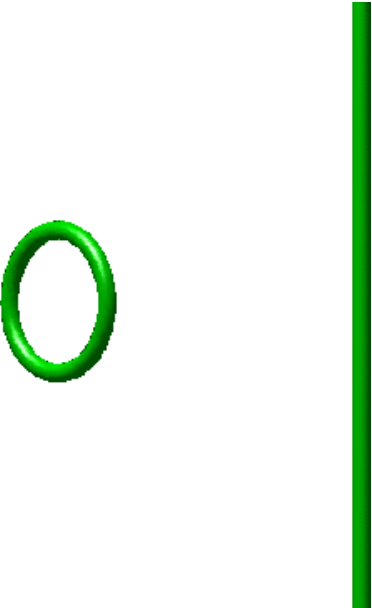}
      \hspace{2cm}
      \includegraphics[scale=0.5]{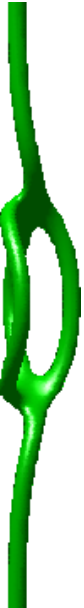} 
      \hspace{2cm}      
      \includegraphics[scale=0.5]{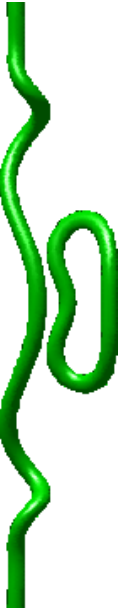}
      \hspace{2cm}      
      \includegraphics[scale=0.5]{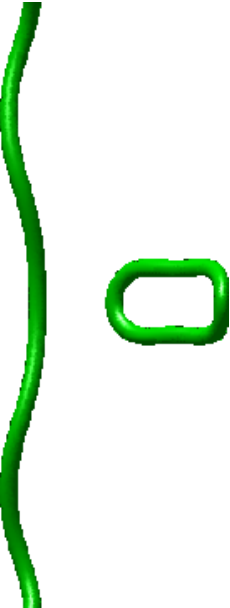}}                 
  \end{minipage}%
  \vspace{0.1cm}
 \begin{minipage}[b]{1.0\textwidth}
    \vspace{0pt}
    \centering
    \subfigure[]{
%      \label{fig:mini:subfig:a}
%      \epsfig{file=figs/Fig_VLdens_comp1.eps,scale=0.014,angle=0.0}}
      \includegraphics[scale=0.5]{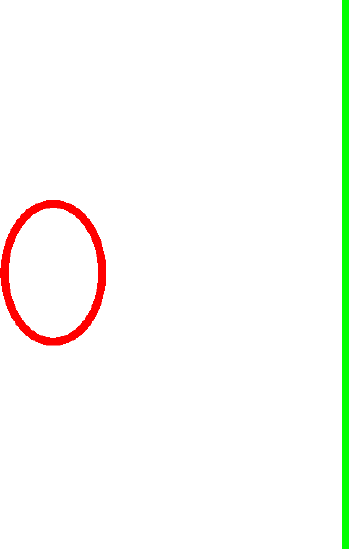}
      \hspace{2cm}      
      \includegraphics[scale=0.5]{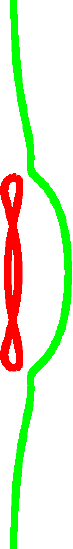}
      \hspace{2cm}        
      \includegraphics[scale=0.5]{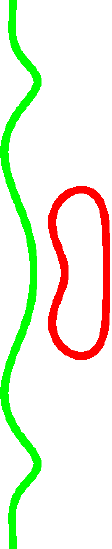}
      \hspace{2cm}      
      \includegraphics[scale=0.5]{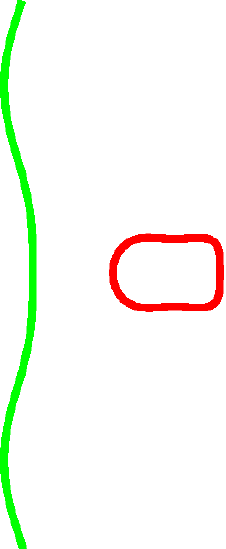}}                
  \end{minipage}
\caption{Time sequence of scattering of a vortex ring by a line vortex computed by integrating the GP equation (\ref{GPadim}) shown at $t=0$, $225$, $250$, $300$: (a) iso-surfaces of the density field at the value $ |\psi|^2=0.3$; (b) Vortex filament representation of topological excitations obtained by tracking zeros of the wavefunction (color figure online).}\label{collision}
\end{figure}

In Fig.\ \ref{collision}(a), we present an example of the scattering of a vortex ring by a line vortex for an initial condition corresponding to $d_y/R = 0$. In general, the dynamics reveal the sequence of events (i) the ring approaches the line vortex with a trajectory that is slightly deviated by the induced velocity field; (ii) if the two objects are sufficiently close to one another, a reconnection takes place and a new ring escapes from the line; (iii) excitations in the form of Kelvin waves are generated both on the line and on the ring. These features are clearly discernible in Fig.\ \ref{collision}(a). 

In order to quantify the properties of the interaction between the two objects, it is useful to evaluate the momentum of each topological excitation. The GP model can be used to evaluate the total energy and momentum of the system arising from the total contribution associated with all excitations present in the flow. However, in the limit when the healing length is very small compared to other characteristic length scales within the flow, it is often possible to approximate the flow by an incompressible system as described in Section 2. In principle, this allows us to attribute momentum and energy to different contributions arising from the Biot-Savart integral for each vortex. In order to do this, we 
need to find the collection of points describing the vortex defects, 
by tracking the vortices in the computed GP wave-function.
To accomplish this, we have developed an algorithm to find a zero of the wave-function in a plane based on the method used by Krstulovic\cite{Krstulovic2012}. By applying this to the $x$-$y$, $x$-$z$, and $y$-$z$ planes of our numerical domain, and appropriately connecting all the points, we are able to separately extract the ring and the line vortex from the wavefunction field as presented in Fig.\ \ref{collision}(b). The results are in agreement with the vortices  identified by the low-density isosurfaces of Fig.\  \ref{collision}(a) which confirms the reliability of our tracking algorithm. 

\begin{figure}[t]
 \begin{minipage}[b]{0.48\textwidth}
    \vspace{0pt}
    \centering
    \subfigure[]{
%      \label{fig:mini:subfig:a}
%      \epsfig{file=figs/Fig_VLdens_comp1.eps,scale=0.014,angle=0.0}}
      \includegraphics[scale=0.22]{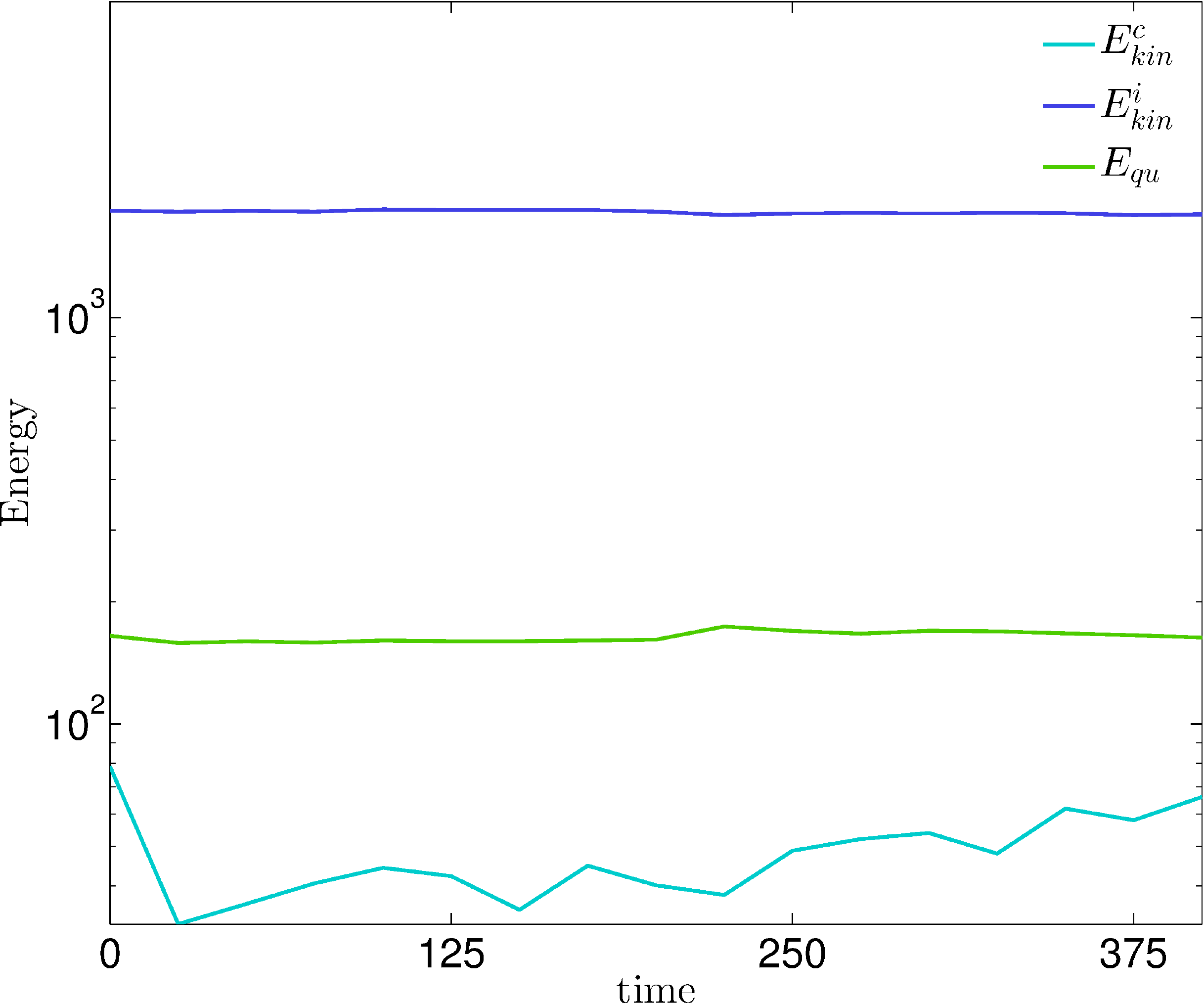}}
  \end{minipage}
  \hspace{0.1cm}
 \begin{minipage}[b]{0.48\textwidth}
    \vspace{0pt} 
    \centering
    \subfigure[]{
%      \label{fig:mini:subfig:a}
%      \epsfig{file=figs/Fig_VLphase_comp1.eps,scale=0.014,angle=0.0}}
      \includegraphics[scale=0.22]{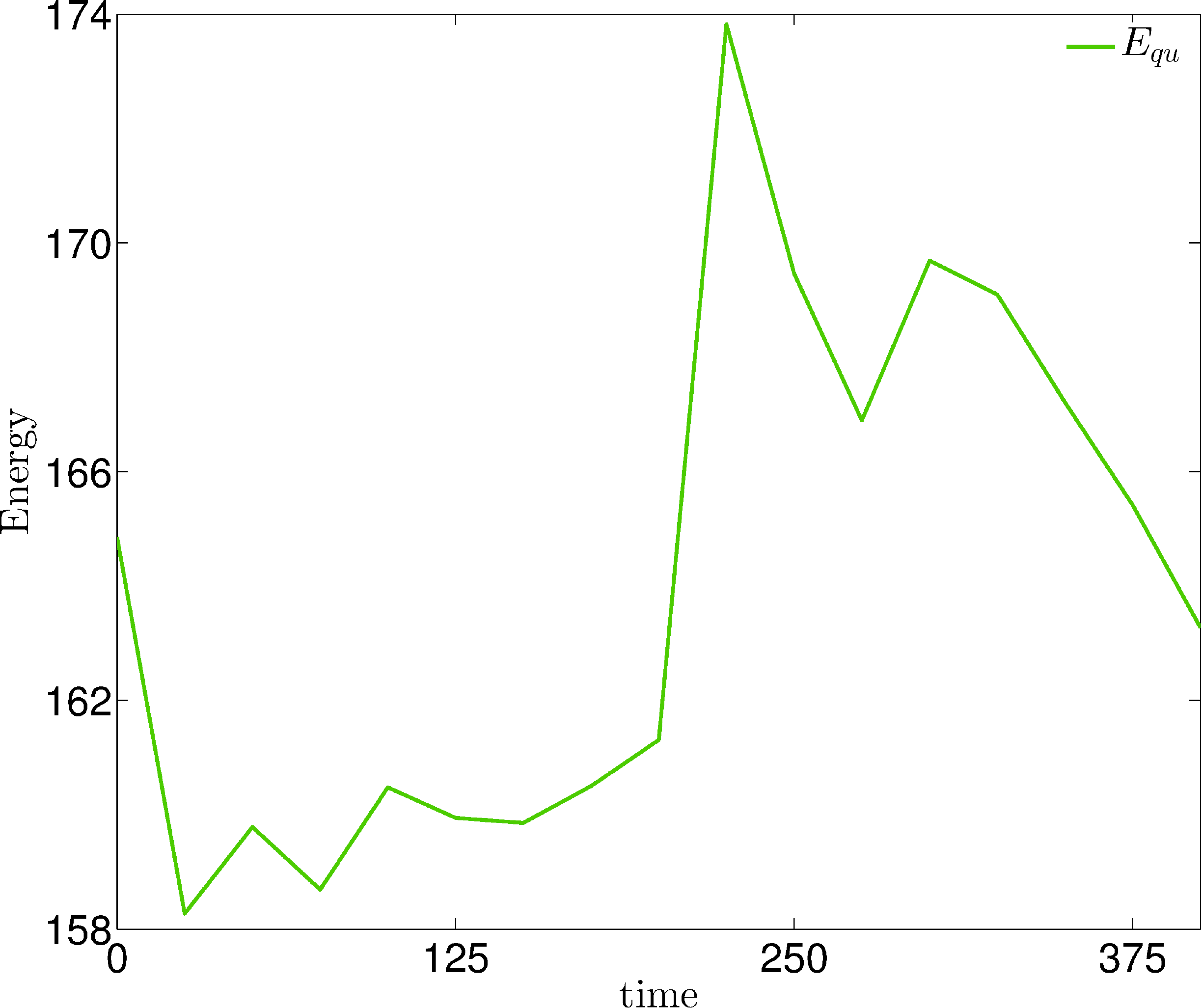}}
  \end{minipage}
\caption{Variation of energy as function of time: (a) Different components of the energy on a semi-logarithmic plot, (b) Variation of quantum energy during a reconnection (color figure online).}\label{veldec}
\end{figure}

For the same run, we evaluated how the different contributions  to the GP energy vary during the scattering. The evolution of the quantum $E_{qu}$ energy, the incompressible kinetic energy $ E_{kin}^i $, and the compressible kinetic energy $E_{kin}^c$ as given in Eqs.\ (\ref{eqn_En_cont}) and (\ref{eqn_Ek_inc_comp}) are plotted in Fig.\ \ref{veldec}(a). The results confirm that the contribution from the compressible kinetic energy is negligible with respect to the incompressible one; it remains two orders of magnitude smaller in comparison to the other components even during reconnections. 
Upon closer inspection of the quantum energy as shown in Fig.\ \ref{veldec}(b), we observe a sharp increase within the time interval between $t=[200,250]$.
This time period coincides with a reconnection event as indicated in Fig.\ \ref{collision}. Despite the substantial increase in the quantum energy during such events, it remains significantly below the incompressible kinetic energy. This verifies that our simulations are performed within a regime where a hydrodynamic approximation is expected to be valid.
For this reason, we will analyse our results in terms of the Biot-Savart or LIA models after applying the vortex tracking algorithm.
In particular, we will focus on how the scattering process depends on the initial ring offset by varying this parameter within the interval $d_y/R \in [-1.5, 1.5]$.  
We point out that the initial distance of the vortex ring from the line vortex is a crucial parameter in the simulation. For the scattering problem, we are strictly interested in the limit of large initial separations $d_x/R$. Given the constraints imposed by our computational domain, we have prescribed the initial distance as $d_x/R=6$. 
To extract the information on the final quantities after the scattering, we 
computed the centre of mass of the ring $ \vv{x}_{CM}^{ring} $ as the average of the positions of the points tracked with the algorithm. Once the ring had reached a threshold distance of $d/R=6$ from the line vortex, the ring-line interaction can be assumed to be small at which point the relevant quantities were computed.

\begin{figure}
\centering
\hspace{-0.8cm}
 \begin{minipage}[b]{0.46\textwidth}
    \vspace{0pt}
    \centering
    \subfigure[]{
%      \label{fig:mini:subfig:a}
%      \epsfig{file=figs/Fig_VLdens_comp1.eps,scale=0.014,angle=0.0}}
      \includegraphics[scale=0.28]{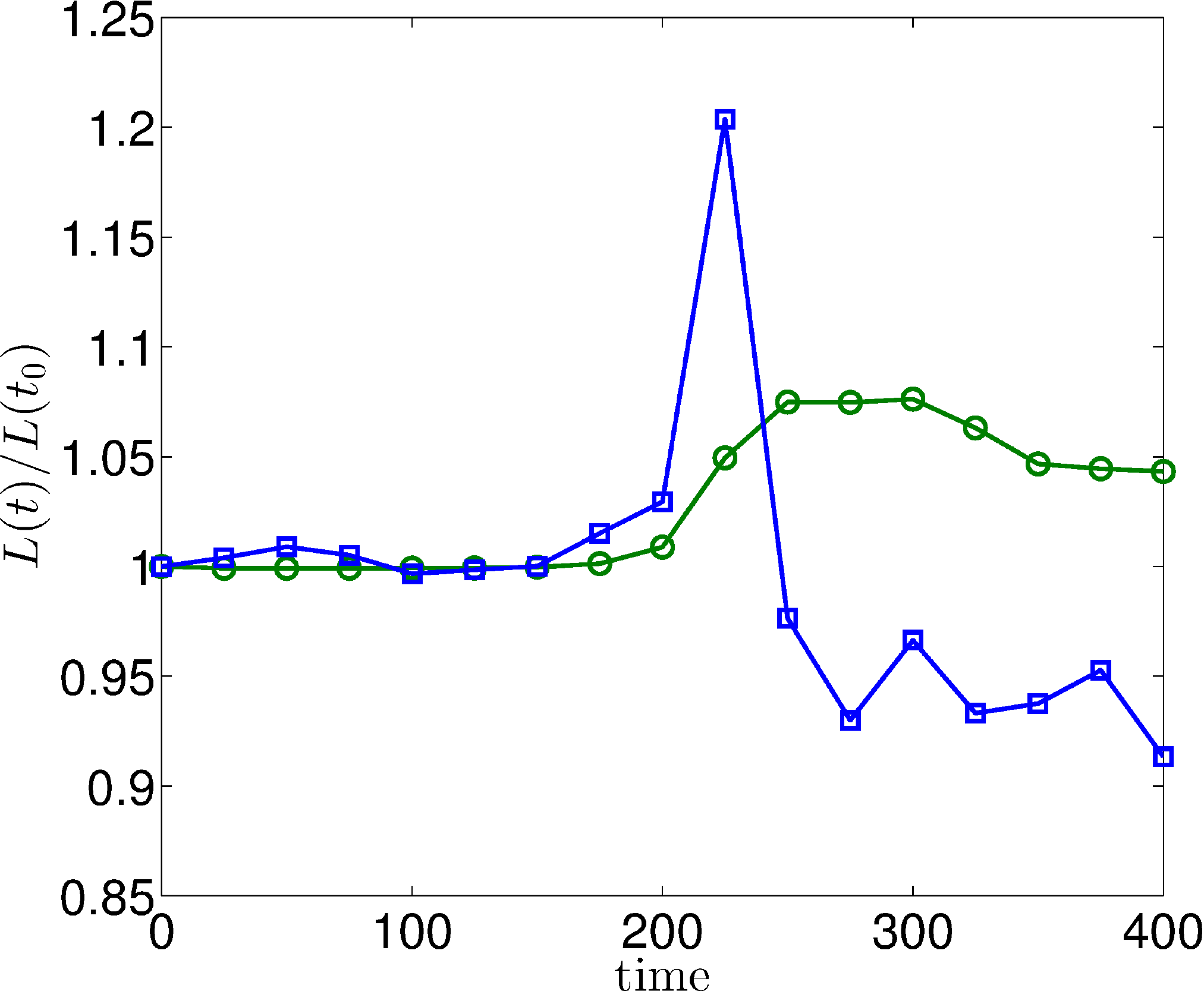}}
  \end{minipage}
  \hspace{0.5cm}
 \begin{minipage}[b]{0.46\textwidth}
    \vspace{0pt} 
    \centering
    \subfigure[]{
%      \label{fig:mini:subfig:a}
%      \epsfig{file=figs/Fig_VLphase_comp1.eps,scale=0.014,angle=0.0}}
      \includegraphics[scale=0.285]{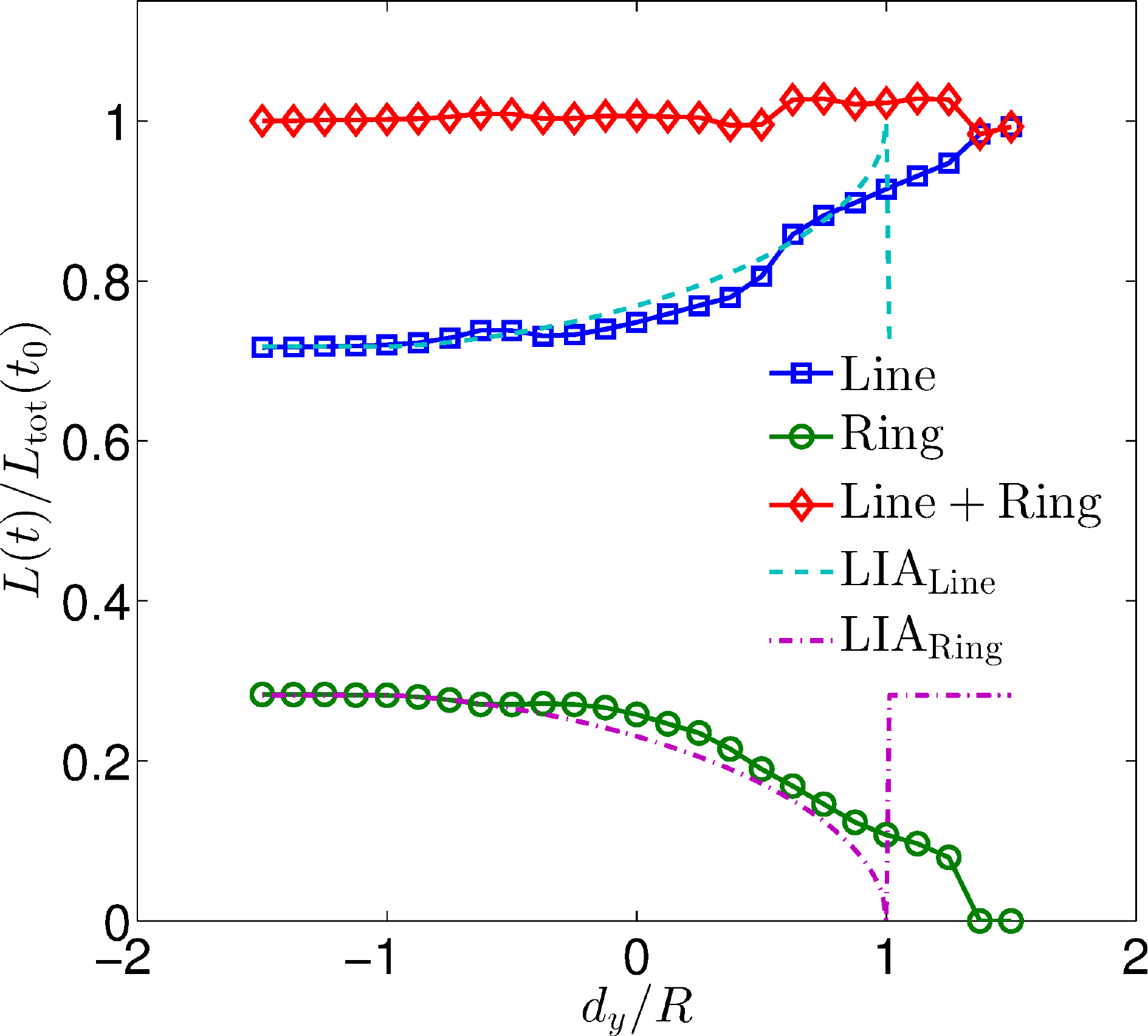}}
  \end{minipage}
\caption{(a) Variation of length of line and ring vortex as a function of time during a reconnection; (b) Final length for the line vortex and the ring normalised by the total initial initial length for different values of the initial offset parameter $d_y/R$. Also shown is the theoretical prediction made with the LIA (color figure online).
\label{graph}}
\end{figure}

We begin by presenting in Fig.\ \ref{graph}(a) results for the variation of the vortex ring and line lengths as a function of time for the case corresponding to $d_y/R =0$, where each length has been normalised with respect to its initial value. We observe that both lengths remain essentially constant up to the point of a reconnection, during which
the line vortex absorbs part of the ring length. At later times both lengths show strong fluctuations due to the presence of propagating Kelvin waves on the two vortices.
In Fig.\ \ref{graph}(b), we evaluated the ratio of the initial and final lengths separately for the vortex ring $L_{\mathrm{ring}}$ and the line vortex $L_{\mathrm{line}}$ as a function of initial ring offset $d_y/R$. Also presented is the total length. For comparison, we have included predictions obtained from the LIA in which the reconnected segments of the line and ring vortices can be obtained from purely  geometric considerations.
The most striking feature of this plot is the asymmetric dependence of the computed line length on initial vortex offset. This is attributed to the severe constraints imposed by the winding around the line vortex and the vortex ring which enforces a very specific change of topology onto the system. Consequently, for our configurations, positive offset values lead to smaller rings being produced whereas negative offset values result in larger rings. Despite the strong nonlocal nature of the line vortex interaction that arises from the Biot-Savart integral upon close approach of the ring to the line vortex, it is interesting that the LIA captures the integrated quantities such as line length quite well away from the reconnections.
Close inspection of Fig.\ \ref{graph}(b) shows that the agreement between the LIA predictions and numerical results is particularly good within the interval $d_y/R \in [-1.5,-0.5]$. To explain this, we note that within the LIA, a reconnection can only occur  if the projection of the initial vortex ring position in the direction along the centreline of the ring crosses the line vortex. This can only occur for $d_y/R \in [-1,1]$. Hence, reconnections are not permitted outside this interval which leads to the stepwise jump at $d_y/R = 1$ seen in Fig.\ \ref{graph}(b). For $d_y/R \in [-1.,-0.5]$, reconnections occur within the LIA but lead to small depletions in the length of the ring. On the other hand, GP simulations show no reconnections within this interval. This is caused by the nonlocal nature of the line vortex interaction in which the ring significantly distorts the line such that it can pass by without reconnecting. For this reason, the agreement between the GP and LIA turns out to be exceptionally good within the interval $d_y/R \in [-1.5,-0.5]$. For $d_y/R > -0.5$, reconnections begin to occur within the GP simulations which results in a slight departure of the calculated final line lengths in comparison to LIA. For values of $d_y/R>1$, the discrepancies are greatest since reconnections still occur in our GP simulations but cease for LIA due to the geometric considerations discussed above.

\begin{figure}
\centering
%\hspace{-0.5cm}
 \begin{minipage}[b]{0.48\textwidth}
    \vspace{0pt}
    \centering
    \subfigure[]{
%      \label{fig:mini:subfig:a}
%      \epsfig{file=figs/Fig_VLdens_comp1.eps,scale=0.014,angle=0.0}}
      \includegraphics[scale=0.2]{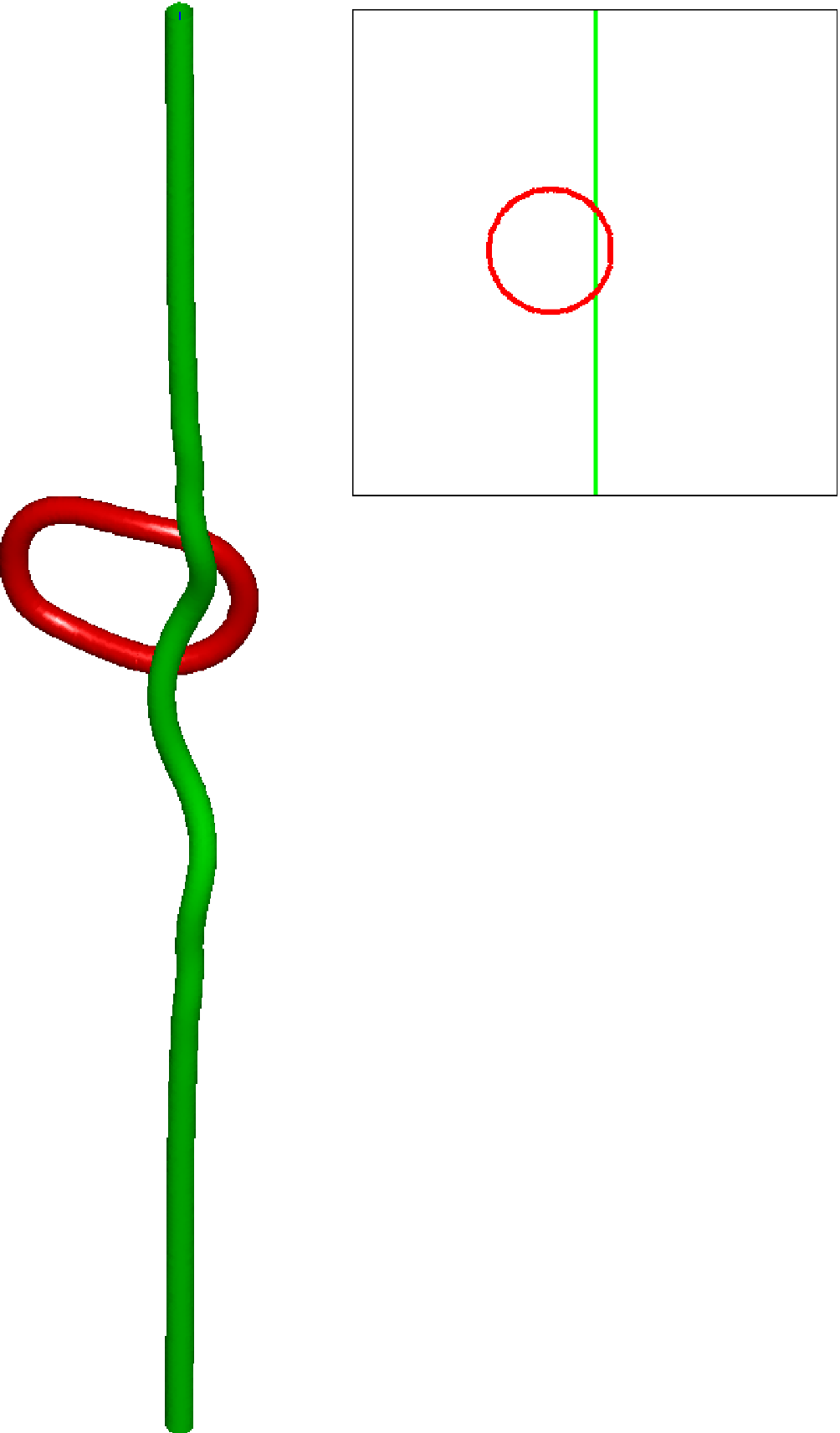}}
  \end{minipage}
  \hspace{0.1cm}
 \begin{minipage}[b]{0.48\textwidth}
    \vspace{0pt} 
    \centering
    \subfigure[]{
%      \label{fig:mini:subfig:a}
%      \epsfig{file=figs/Fig_VLphase_comp1.eps,scale=0.014,angle=0.0}}
      \includegraphics[scale=0.2]{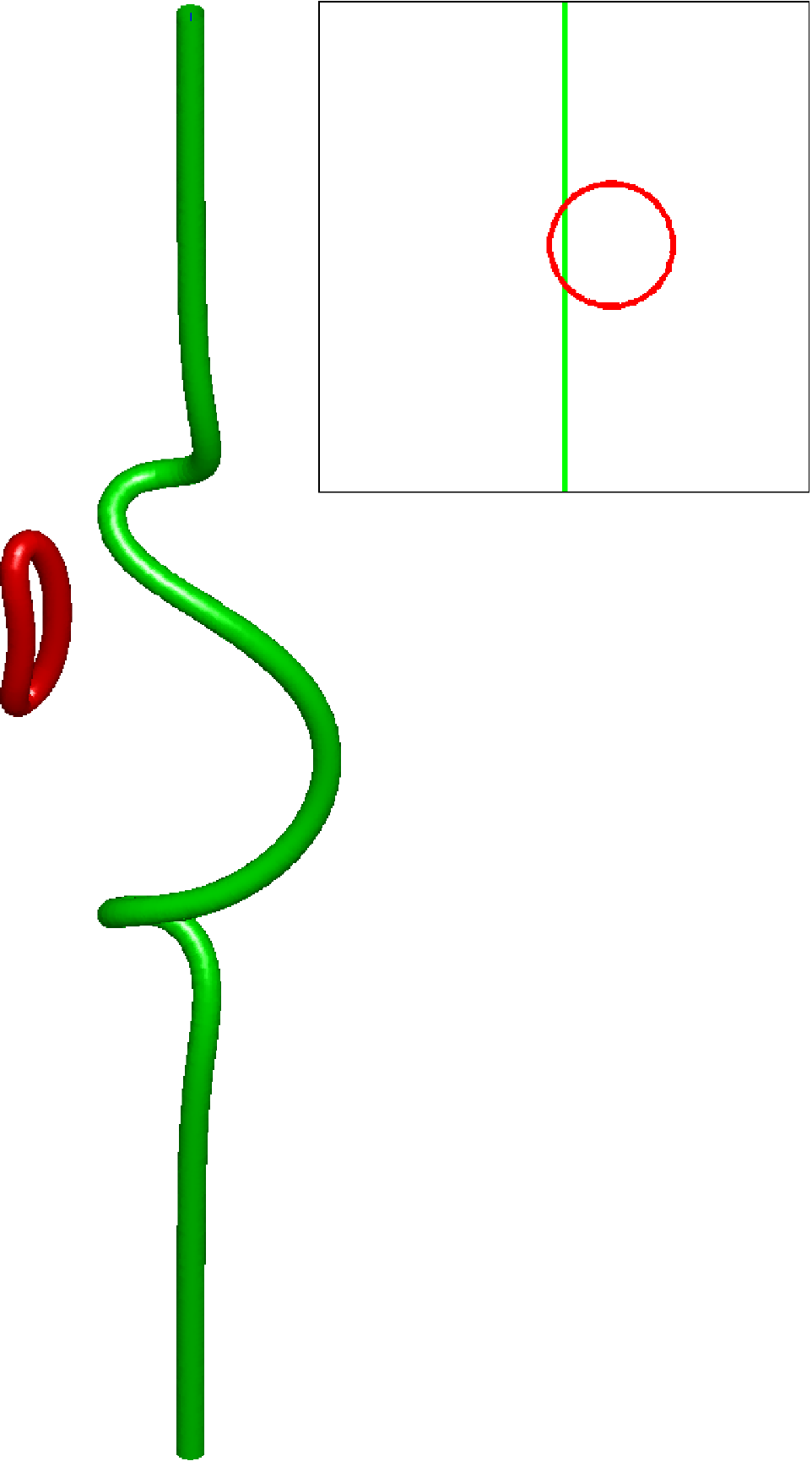}}
  \end{minipage}
\caption{Asymmetric behaviour for two collisions having two opposite starting offset positions ($d_y/R$). Different colours are used for the vortex line and ring for ease of visualisation; (a) $d_y/R=-0.75 $, (b) $d_y/R=0.75 $. The initial vortex ring positions relative to the initial location of the line vortex are shown in the insets (color figure online). 
\label{Asyline}}
\end{figure}

\begin{figure}
\centering
%\hspace{-0.5cm}
 \begin{minipage}[b]{0.48\textwidth}
    \vspace{0pt}
    \centering
    \subfigure[]{
%      \label{fig:mini:subfig:a}
%      \epsfig{file=figs/Fig_VLdens_comp1.eps,scale=0.014,angle=0.0}}
      \includegraphics[scale=0.25]{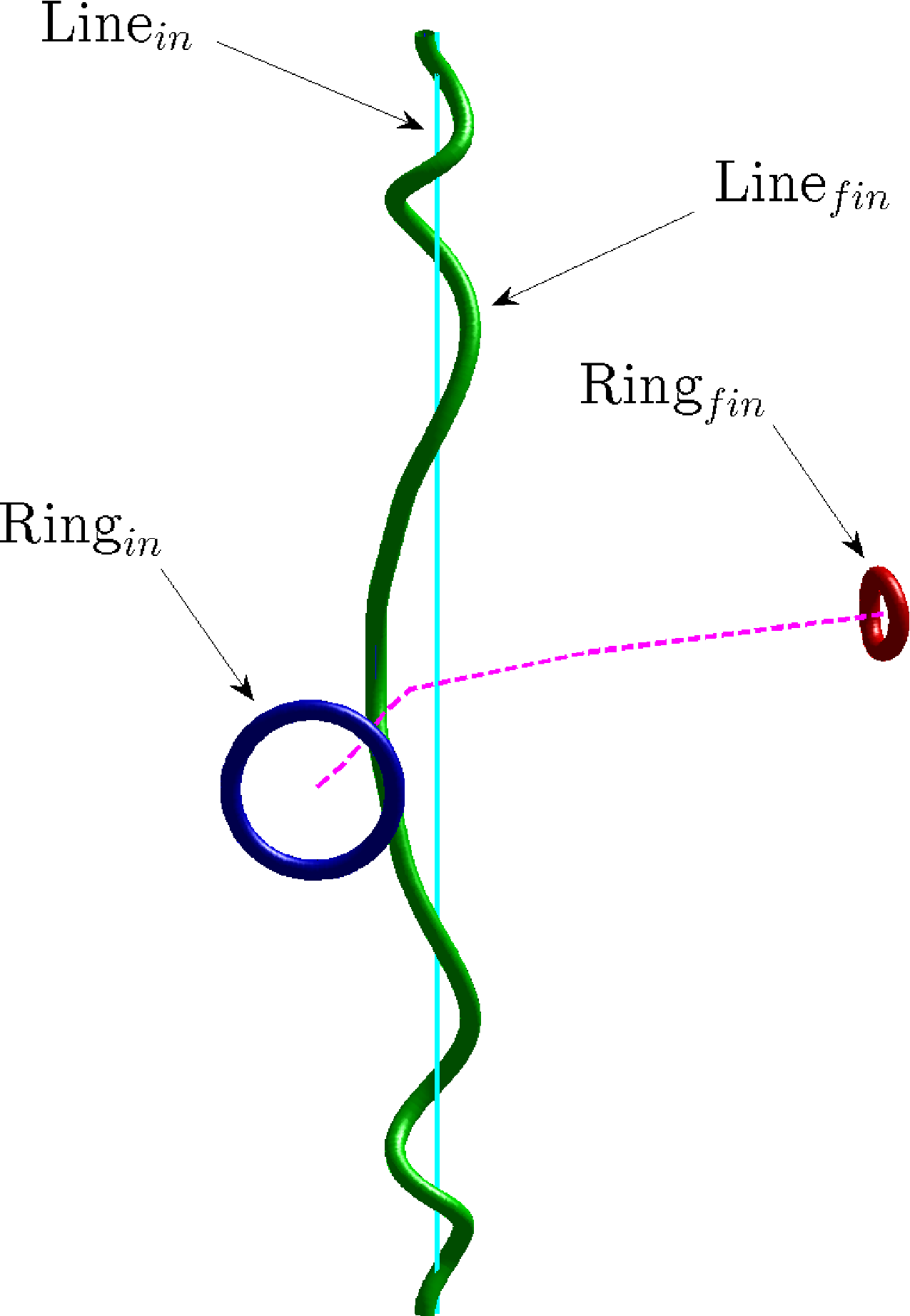}}
%      \vspace{0.1cm}
  \end{minipage}
  \hspace{0.1cm}
 \begin{minipage}[b]{0.48\textwidth}
    \vspace{0pt} 
    \centering
    \subfigure[]{
%      \label{fig:mini:subfig:a}
%      \epsfig{file=figs/Fig_VLphase_comp1.eps,scale=0.014,angle=0.0}}
      \includegraphics[scale=0.2]{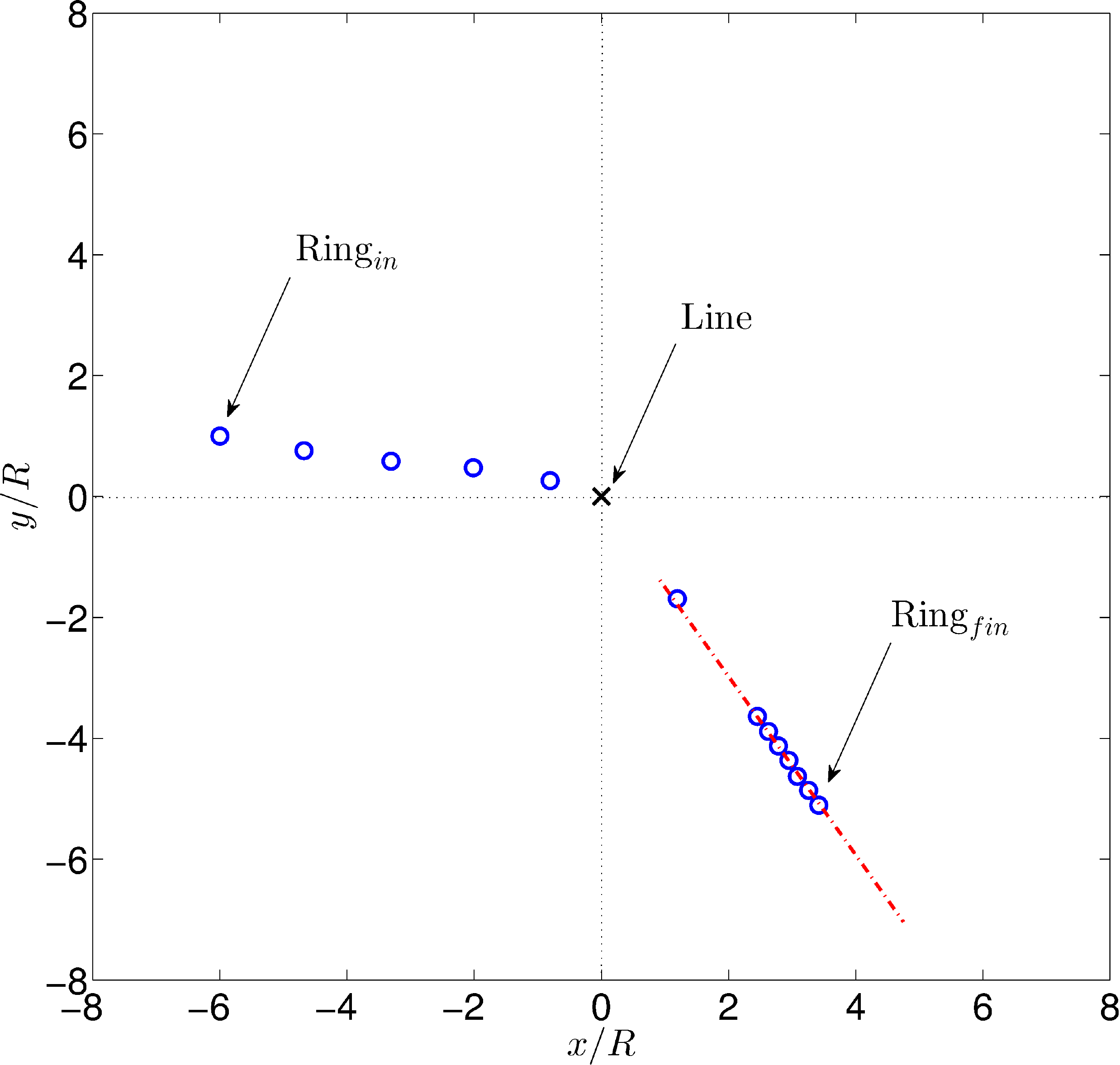}}
  \end{minipage}
\caption{(a) Illustration of the scattering of a ring by a line resulting in line length loss and deflection of the ring. (b) Trajectory of a ring on the $x$-$y$ plane during an encounter with a line vortex. The vortex ring positions are plotted at the times, $t=0$, $50$, $100$, $150$, $200$, $250$, $300$, $306$, $312$, $318$, $324$, $330$, $336$ (color figure online).
\label{fig:CM}}
\end{figure}

As a clear illustration of the asymmetry seen in our simulations,  we present in Fig.\ \ref{Asyline} plots  of the vortex line and ring positions following a reconnection event for two opposite values of the ring offset parameter. These plots were obtained by extracting the vortex positions using our tracking algorithm to elucidate the differences seen for the 
two different values of the initial offset corresponding to ($d_y/R=-0.75$) and to ($d_y/R=0.75$). Due to the conservation of the circulation, reconnections always lead to  the new ring being located to the {\it left} in our figures. Thus the emitted rings have very different size following their encounter with the line vortex.
Under the LIA, smaller circular rings are known to travel faster than larger ones. We can, therefore, infer the velocities of the rings from knowledge of their size. However, we are in the position to evaluate their velocities directly from the variation of the ring centre of mass $ \vv{x}_{CM}^r $ with time. 
%Figure \ref{fig:CM} presents the key important features of a single numerical realization associated with a vortex ring simulation. 
Figure \ref{fig:CM}(a) provides an illustration of how an initially circular vortex ring scatters with a straight line vortex. The computed trajectory of the ring is also included to illustrate the deflection that a ring experiences as it encounters the line vortex.
Since the linear momentum of the initial condition has no $z$-component (the single line carries null linear momentum and the axis of the ring is by construction aligned along the $x$-coordinate direction), we expect the motion of the ring to be purely on the $x$-$y$ plane. Therefore, in Fig.\ \ref{fig:CM}(b) we have plotted the position of the ring at different times on the $x$-$y$ plane. In this plot, lengths are measured in units of the initial ring radius $R$. The results indicate that in the latter stages of the simulation, the ring moves with almost linear velocity.

\begin{figure}
\centering
\hspace{-0.5cm}
 \begin{minipage}[b]{0.48\textwidth}
    \vspace{0pt}
    \centering
    \subfigure[]{
%      \label{fig:mini:subfig:a}
%      \epsfig{file=figs/Fig_VLdens_comp1.eps,scale=0.014,angle=0.0}}
      \includegraphics[scale=0.24]{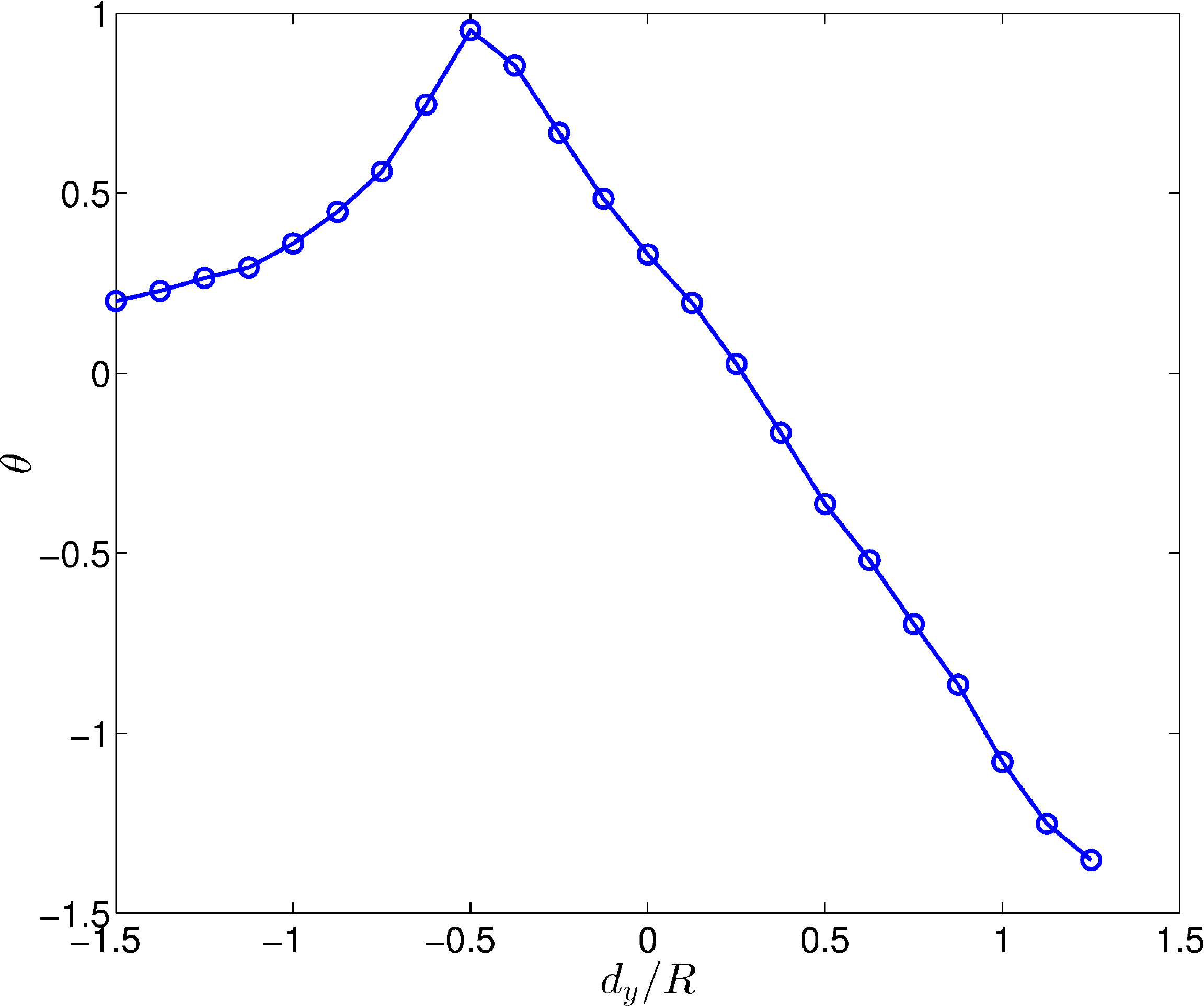}}
%      \vspace{0.1cm}
  \end{minipage}
  \hspace{0.2cm}
 \begin{minipage}[b]{0.48\textwidth}
    \vspace{0pt} 
    \centering
    \subfigure[]{
%      \label{fig:mini:subfig:a}
%      \epsfig{file=figs/Fig_VLphase_comp1.eps,scale=0.014,angle=0.0}}
      \includegraphics[scale=0.24]{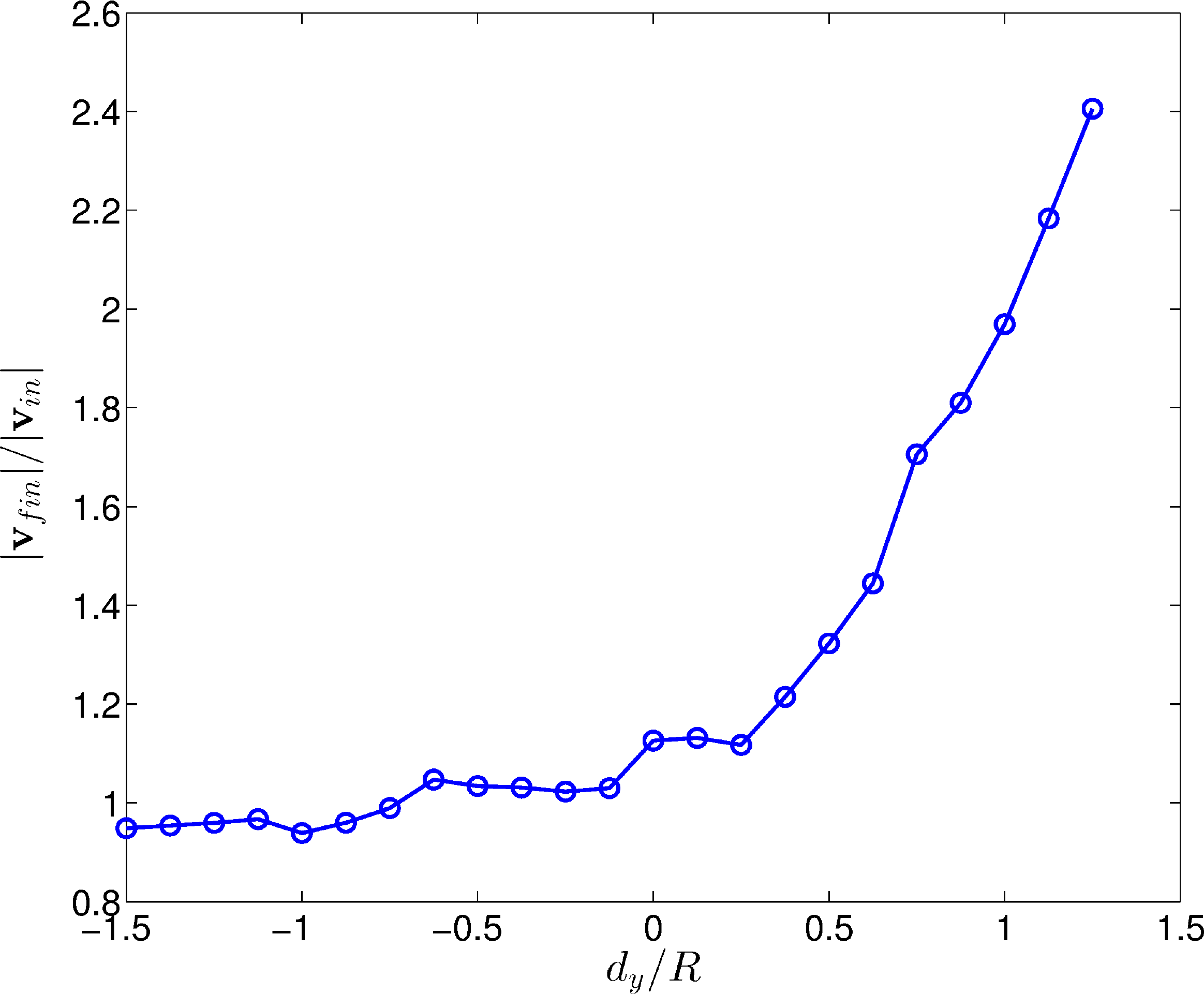}}
  \end{minipage}
\caption{(a) Variation of the deflection angle $\theta$ as a function of the initial offset $d_y/R$; (b) Variation of the final speed of the vortex ring as a function the initial offset $d_y/R$ (color figure online).
\label{anglevelocity}}
\end{figure}

To simplify the analysis, we can assume that, at the initial and final stages, the motion of the ring is essentially uniform as the interaction with the line vortex is weak. By fitting the trajectory of the ring given by its centre of mass with a straight line, we are able to evaluate the characteristic deflection angle $\theta=\theta_{fin}-\theta_{in}$ as well as the variation in the velocity magnitude $|\vv{v}_{fin}|/|\vv{v}_{in}|$.
Figures \ref{anglevelocity}(a) and \ref{anglevelocity}(b) show the behaviour of these two parameters versus the initial offset value $d_y/R$, respectively.
The deflection angle depends strongly on whether or not a reconnection occurs. 
As discussed previously, reconnections take place for offset values greater than $d_y/R=-0.5$, resulting in a drastic change in the behaviour of the deflection angle.
On the other hand, the variation in the velocity magnitude grows as a smooth and essentially monotonically increasing function of the initial offset. As expected, smaller rings that are produced at positive values of the offset, have very large velocities. It is important to note that when smaller rings are produced,  a larger fraction of the energy is transferred into large amplitude excitations along the line vortex. These subsequently could interact as soliton like excitations\cite{Hasimoto1972}. On the other hand, the emission of large rings imparts small amplitude Kelvin-wave excitations onto the line vortex.

\section{Conclusions}
We have studied the scattering of vortex rings by a line vortex using the GP equation. In particular, we have identified the effect of varying the initial offset parameter of the vortex ring on its subsequent scattering properties. Our results have revealed that a strong asymmetry is seen in the scattering of vortex rings by a line vortex due to the severe topological constraints imposed on the system. By using a vortex tracking algorithm, we have evaluated how the length of the ring and line vary after scattering as a function of the initial offset parameter. We showed that predictions made by LIA based upon simple geometric considerations lead to reasonably good agreement with results based on full GP simulations. Moreover, we find that reconnections resulting in the emission of smaller vortex rings that move at higher velocities result in strong deflections in the trajectory of the ring whilst the emission of larger rings results in weak deflections. 

The production of smaller rings during scattering corresponds to the deposition of a larger fraction of the energy from the ring onto the line. This leads to large amplitude Kelvin waves that can behave more like Hasimoto solitons\cite{Hasimoto1972}  propagating along a vortex. On the other hand, larger loops correspond to small amplitude Kelvin waves being produced on the line vortex. These observations have direct relevance to our understanding of the cross-over range of scales in superfluid turbulence where the emission of vortex rings due to the direct energy cascade as discussed by Svistunov\cite{Svistunov1995} or due to breather excitations as reported by Salman\cite{Salman2013} turns out to be very important. Indeed, the production of large amplitude Kelvin waves can result in the emission of further rings following self-reconnections on a vortex line. These findings raise the question of how transparent is a turbulent tangle to vortex rings that can be emitted within the cross-over range as this can have direct relevance to the question of how energy is dissipated within the ultra-low temperature regime of turbulence.

\bibliography{ref}{}
\bibliographystyle{plain}

\end{document}